\documentclass[prd,twocolumn,preprintnumbers,nofootinbib,amsmath,amssymb]{revtex4}

\usepackage{graphicx}
\usepackage{dcolumn}
\usepackage{bm}

\begin{document}

\preprint{IFUP-TH/2005-11}

\title{Color confinement and dual superconductivity of the vacuum. IV.}

\author{M. D'Elia$^{1}$, A. Di Giacomo$^{2}$, B. Lucini$^{3}$, G. Paffuti$^{2}$, C. Pica$^{2}$}

\affiliation{$^1$  Dipartimento di Fisica dell'Universit\`a di Genova and INFN, Sezione di Genova, Via Dodecaneso 33, I-16146 Genova, Italy}
\affiliation{$^2$ Dipartimento di Fisica dell'Universit\`a and INFN sezione di Pisa, L.go Pontecorvo 3 Ed. C, I-56127 Pisa, Italy}
\affiliation{$^3$ Institute for Theoretical Physics, ETH Z\"urich, CH-8093 Z\"urich, Switzerland}

\date{\today}

\begin{abstract}
A scaling analysis is made of the order parameter describing monopole
condensation at the deconfining transition of $N_f=2$ QCD around the
chiral point. In accordance with scaling properties of the specific
heat, studied in a previous paper, scaling is consistent with a first
order transition. The status of dual superconductivity of the vacuum
as a mechanism of color confinement is reviewed.
\end{abstract}

\maketitle

\section{\label{SECT1}Introduction}

This paper is the fourth of a
series~\cite{PaperI,PaperII,PaperIII}\footnote{In the following we
shall refer to the previous papers as I, II, III.} on a research
program aiming at developing and testing on the lattice the working
hypothesis that confinement in QCD is produced by dual
superconductivity of the vacuum, i.e. by condensation of magnetic
charges.

Such a mechanism implies that the deconfining transition is an
order--disorder phase transition, i.e. a true phase transition,
described by an order parameter.

Another basic underlying idea is
duality~\cite{Kramers:1941kn,Kadanoff:1970kz}: the magnetic
excitations which are expected to condense are non local in terms of
the fields. They should become local in the dual description, and the
strong coupling regime of the confined phase should be mapped by
duality in the weak coupling regime for dual fields.

Magnetic charges exist in gauge
theories~\cite{Dirac:1931kp,'tHooft:1974qc,Polyakov:1974ek} and are
nicely described in a formulation of the theory based on parallel
transport, as the Wilson formulation is ~\cite{Wilson:1974sk}.

The prototype theory is the lattice $U(1)$ gauge theory in 4d. In the
Wilson's formulation this theory shows a deconfining transition at
$\beta\equiv 2/g^2 \simeq 1.01$, from a strong coupling phase in which
static charges are confined, to a Coulomb phase. Dual magnetic
excitations can be defined and with them an order parameter which is
the \textit{vev} $\langle\mu\rangle$ of a magnetically charged
operator $\mu$. This \textit{vev} is different from zero in the
confined phase, and strictly zero in the deconfined phase. This
operator is defined by its correlation
functions~\cite{Frohlich:1986sz,Frohlich:1987er,Frohlich:1982gf},
exactly like the kink's correlators in the Ising model of
Ref.~\cite{Kadanoff:1970kz}. It can however be explicitly defined as a
non local operator in terms of the gauge
fields~\cite{Marino:1981we,DiGiacomo:1997sm} in the same way as the
kink creation operator can be explicitly written in terms of the spin 
 variables~\cite{Carmona:2000eu}.

The operator $\mu$ is a gauge invariant Dirac
like~\cite{Frohlich:1986sz,Frohlich:1987er,Frohlich:1982gf} magnetic
charged operator, non local but obeying cluster property, and a
rigorous proof exists that it is the correct order parameter, both for
the Villain~\cite{Frohlich:1987er} and for the
Wilson~\cite{Cirigliano:1997zq} actions.  In the presence of electric
charges, however, there are problems for the existence of the duality
transformation~\cite{Frohlich:2000zp}. We will come back below in
Sect.~\ref{SECT2}.

The translation of the above formalism into non abelian gauge theories
is non trivial.  The original proposal to define and expose
monopoles~\cite{'tHooft:1981ht} is based on a procedure known as
abelian projection. A gauge is fixed, e.g. by diagonalizing any
observable in the adjoint representation, a residual $U(1)^{N-1}$
gauge invariance is left and the monopoles are defined as monopoles of
each of these $U(1)$s, as for the abelian case.  They show up as
singularities of the gauge transformation which brings to the abelian
projected gauge, in the sites in which two eigenvalues of the chosen
observable coincide.

In this way the number and the location of monopoles in a given field
configuration depends on the abelian projection, i.e. on the choice of
the gauge, and there is a functional infinity of ways to do it.

Therefore if one pretends that the monopoles detected on the lattice
configurations are the relevant degrees of freedom for confinement
(monopole dominance~\cite{Suzuki:1989gp}), then one has to select the
abelian projection which defines them. The usual choice is the so
called maximum abelian projection, by which a gauge is selected
maximizing the role of the residual abelian degrees of freedom. In
some literature this choice is called the abelian projection
\textit{tout-court}. According to this approach the monopoles in
this projection are the effective degrees of freedom for confinement.

An alternative attitude is to look at the change of symmetries at the
transition, postponing the identification of the relevant excitations,
which is a far more difficult problem. This amounts to define and test
an order parameter $\langle\mu\rangle$ for monopole condensate.

The lattice version of the operator $\mu$ has been constructed
in~I,~II. $\mu(\vec x, t)$ creates a monopole in a given abelian
projection at the site $(\vec x,t)$. $\mu$ is color gauge invariant,
magnetic $U(1)$ gauge invariant~\cite{Frohlich:1982gf}, but of course
is abelian projection dependent.

In~III it was shown numerically that the order parameter
$\langle\mu\rangle$ is abelian projection independent. Arguments were
given in Ref.~\cite{DiGiacomo:2002pe} justifying theoretically this
observation.  Below we will present a complete argument showing that
the statement $\langle\mu\rangle\neq0$ or $\langle\mu\rangle=0$ is
abelian projection independent.  Any $\mu$ creates a monopole in all
abelian projections. The recent observation of the existence of
abelian Abrikosov flux tubes even without fixing the gauge fully
supports this view~\cite{Suzuki:2004uz}.

Creating a monopole is a better defined operation than detecting it;
the same is true also for $Z_n$ vortices~\cite{vort1}. Their creation
is gauge independent; their detection depends on the choice of the
gauge.

In summary $\langle\mu\rangle$ proves to be a good order parameter for
the quenched theory: it is gauge invariant, abelian projection
independent, strictly zero in the deconfined phase in the infinite
volume limit~\cite{PaperI,PaperII,D'Elia:2003xn}, non zero in the
confined phase and its scaling with volume is consistent with
the critical indexes of the transition.

On this basis $\langle\mu\rangle$ can be tentatively used as an order
parameter for confinement also in full QCD. This would fill a gap:
indeed in full QCD no $Z_3$ symmetry exists, the chiral phase
transition is defined at $m_q=0$, but an order parameter for
confinement is missing.

In Ref.~\cite{Carmona:2002ty,Carmona:2002ye,Carmona:2002yg} it was
proved numerically that, for $N_f=2$, $\langle\mu\rangle\neq0$ below
the ``transition line'' and $\langle\mu\rangle=0$ above it.  In this
paper we discuss a finite size scaling analysis of the order parameter
for $N_f=2$ QCD.  We find that it is consistent with a first order
phase transition. The same indication comes from the analysis of the
specific heat and of the chiral order parameter at the chiral
transition~\cite{CHIPAPER}.

In Sect.~\ref{SECT2} we review the status of the order parameter
$\langle\mu\rangle$: its independence on the abelian projection, the
difficulties related to the coexistence in the theory of electric and
magnetic charges.

In Sect.~\ref{SECT3} we present the numerical results for $N_f=2$ QCD
and a scling analysis of the order of the deconfinement phase
transition.

Sect.~\ref{SECT4} contains the conclusions and perspectives.

\section{\label{SECT2}The (dis)order parameter $\langle\mu\rangle$}
The operator $\mu^a$ ($a=1,\ldots,N_c-1$) which creates a monopole of
the species $a$ in a given abelian projection is defined in the
continuum as~\cite{DiGiacomo:2003ee}
\begin{multline}
\mu^a (\vec x, t) =  \\ 
 \exp \left[ i \int\!\! d\vec y \,\,{\rm Tr} \{ \phi^a(\vec y, t) \vec E (\vec y, t)\} \vec b_\perp (\vec x - \vec y) \right] \label{MUA} \, .
\end{multline}
Here 
\begin{itemize}
\item [-] $\vec b_\perp (\vec x - \vec y)$ is the vector potential produced by a Dirac $U(1)$ monopole sitting at $\vec x$ in the transverse gauge:
\begin{equation}
\vec\nabla\cdot \vec b_\perp = 0 \;\; ; \;\; \vec\nabla \wedge \vec b_\perp (\vec r) = \frac{2\pi}{g}\frac {\vec r}{r^3} + \text{Dirac String} \nonumber \, ;
\end{equation}
\item [-] $\phi^a$ transforms in the adjoint representation and has the form
\begin{equation}
\phi^a(x) = U^\dagger (x) \phi^a_{diag} U(x) \label{PHI}
\end{equation}
with
\begin{equation}
\phi^a_{diag} = \mathrm{diag}\, \Big(\overbrace{\frac{N_c-a}{N_c},\ldots,\frac{N_c-a}{N_c}}^{a},\overbrace{-\frac{\vphantom{N_c}a}{N_c},\ldots,-\frac{a}{N_c}}^{N_c-a} \Big) \nonumber \, .
\end{equation}
The gauge transformation $U$ identifies the representation in which $\phi^a(x)$ is diagonal, i.e. the abelian projection. Of course $\mu^a$ depends on $U$, i.e. on the abelian projection.
\end{itemize}

In the abelian projected gauge $\phi^a=\phi^a_{diag}$, which is $x$
independent, the longitudinal part of $\vec E$ in Eq.~\ref{MUA}
cancels in the convolution with $\vec b_\perp$, and only the diagonal
part of $\vec E$ contributes to the trace on color indexes.  If we
define the matrices $T^a$ ($a=1,\ldots,N_c-1$) as
\begin{equation}
T^a \equiv \mathrm{diag} \big(\underbrace{0,\ldots,0,\overset{a}{1},\overset{a+1}{-1},0,\dots,0}_{N_c} \big) \nonumber
\end{equation}
these matrices are a complete basis for traceless diagonal matrices. Therefore
\begin{equation}
\vec E_{\perp\,\,diag} = \sum \vec E^a_{\perp\,\,diag} T^a \nonumber
\end{equation}
and since
\begin{equation}
{\rm Tr}\{\phi^a_{diag}T^b\} = \delta^{ab} \nonumber
\end{equation}
in the abelian projected gauge
\begin{equation}
\mu^a (\vec x, t) =  \exp \left[ i \int\!\! d\vec y \,\,\vec E^a_{\perp\,\,diag} (\vec y, t) \vec b_\perp (\vec x - \vec y) \right] \nonumber \, .
\end{equation}
$\vec E^a_{\perp\,\,diag}$ is the conjugate momentum to $\vec A^a_\perp$, so that in the Schr\"odinger representation
\begin{equation}
\mu^a (\vec x, t)\,\, \lvert \vec A^a_\perp (\vec x,t) \rangle = \lvert \vec A^a_\perp (\vec x,t) + \vec b_\perp (\vec x - \vec y)\rangle \nonumber
\end{equation}
$\mu^a$ creates a Dirac monopole in the residual $U(1)$ gauge field in
the direction of $T^a$.

On the other hand it can be
proved~\cite{'tHooft:1981ht,PaperI,PaperII} that the 't~Hooft tensor
corresponding to a scalar field $\phi$ in the adjoint representation
\begin{equation}
F_{\mu\nu} \equiv {\rm Tr} \left\{ \phi G_{\mu\nu} - \frac{i}{g} \phi \left[D_\mu\phi , D_\nu\phi \right] \right\} \label{FMUNU1}
\end{equation}
reduces to an abelian form if and only if $\phi$ has the form $\phi^a$
of Eq.~\ref{PHI}.  In that case indeed Eq.~\ref{FMUNU1} becomes
\begin{equation}
F^a_{\mu\nu} (x) = {\rm Tr} \left\{ \partial_\mu ( \phi^a A_\nu) - \partial_\nu ( \phi^a A_\mu) - \frac{i}{g} \phi^a \left[\partial_\mu\phi^a , \partial_\nu\phi^a \right]\right\} \nonumber \, ;
\end{equation}
the quadratic terms in $A_\mu A_\nu$ in the definition of $F_{\mu\nu}$
cancel and
\begin{equation}
\partial_\mu F^{\star a}_{\mu\nu} \equiv \partial_\mu \left( \frac{1}{2} \epsilon_{\mu\nu\rho\sigma} F^a_{\rho\sigma} \right) = 0 \nonumber \, ,
\end{equation}
i.e. the Bianchi identities are obeyed except for singularities.  In
the abelian projected gauge $\partial_\mu\phi^a = 0$, $(A_\mu)_{diag}
= \sum A_\mu^a T^a$ and
\begin{equation}
F^a_{\mu\nu} = \partial_\mu A^a_\nu - \partial_\nu A^a_\mu \nonumber \, .
\end{equation}
For the simple case of $SU(2)$ $a$ only assumes the value $a=1$,
\begin{equation}
\phi^1_{diag}=\frac{1}{2}
\begin{pmatrix} 
1& 0\\ 
0& -1 
\end{pmatrix}
=\frac{1}{2}\sigma_3 \;\; ; \;\; \phi^1 (x) = U^\dagger(x) \frac{1}{2} \sigma_3 U(x) \nonumber \, .
\end{equation}
For $SU(2)$ any hermitian field $\phi(x)$ in the adjoint
representation can be put in the form $\phi(x) =
U^\dagger(x)\frac{1}{2}\sigma_3 U(x)$; meaning that an abelian
projection is obtained by diagonalizing $\phi(x)$.  Monopoles will
appear when the gauge transformation $U(x)$ is singular, and this
happens when $|\phi(x)|=0$, i.e. when two eigenvalues of $\phi(x)$
coincide~\cite{'tHooft:1981ht}.

For generic $SU(N)$ any hermitian field $\phi(x)$ in the adjoint
representation can be written as
\begin{equation}
\phi(x) = \sum c_a(x)  U^\dagger (x) \phi^a_{diag} U(x) \nonumber
\end{equation}
with $U(x)$ the gauge transformation which diagonalizes $\phi(x)$ and
orders the eigenvalues (for instance in a decreasing order for their
imaginary parts). $U(x)$ will be singular at points where any of the
coefficients $c_a(x)$ is zero. In these points two eigenvalues of
$\phi$ coincide. Indeed from the definition of $\phi^a(x)$, in the
abelian projected gauge $\phi^a =
\phi^a_{diag}$ and $\phi_{ii}(x)-\phi_{(i+1)(i+1)}(x) = c_i(x)$.  The
matrices $\phi^a_{diag}$ play in $SU(N)$ the analogous role as
$\sigma_3/2$ in $SU(2)$.

One can define magnetic currents
\begin{equation}
j^a_\mu(x)\equiv\partial_\mu F^{\star a}_{\mu\nu} \nonumber \, .
\end{equation}
Those currents are zero everywhere, because of Bianchi identities,
except at singular points where monopoles are located. In any case
\begin{equation}
\partial_\mu j^a_\mu = 0 \nonumber
\end{equation}
defining a magnetic $U(1)$ symmetry, and conservation of magnetic
charge. If this symmetry is realized \textit{\`a la} Wigner, i.e. if
the vacuum has definite magnetic charge, the system is normal. The
vacuum is a magnetic superconductor if it does not have a definite
magnetic charge, but is a superposition of states with different
magnetic charges.  $\langle\mu^a\rangle \neq 0$ can signal the second
possibility. In the normal state $\langle\mu^a\rangle = 0$

The lattice version of $\mu^a$ has been constructed in~I~and~II, and
is equally well defined in the presence of dynamical quarks. It was
numerically shown in~III that being $\langle\mu^a\rangle = 0$ or
$\langle\mu^a\rangle \neq 0$ is a property independent of the choice
of the abelian projection, and is shared by the version in which
$\phi^a$ is put diagonal configuration by configuration, which is kind
of an average on infinitely many abelian projections.  The same thing
results from comparison of measurements of $\langle\mu^a\rangle$ in
different abelian projections with measurement made by use of
Scr\"odinger functional~\cite{Cea:1996ff} as in
Ref.~\cite{Cea:2001an}.  More recently flux tubes were observed in
$SU(2)$ gauge configurations in the $\sigma_3$ direction without
fixing the abelian projection~\cite{Suzuki:2004uz}.

In fact, if the density of monopoles is finite, i.e. if in whatever
abelian projection there exists a finite number of them per unit
volume, then the gauge transformation $V(x)$ which connects two
abelian projections is continuos everywhere except at a finite number
of points, where monopoles are located.

Creating a monopole with $\mu^a(x)$ will add a singularity say in one
of the abelian projections: the probability that $x$ coincides with
the position of a monopole already present will be zero, and there
will be a neighbourhood of $x$ in which $V$ is regular. Adding a
singularity in one of the abelian projections amounts to add a new
singularity also in the others. $\mu^a$ creates a monopole in all
abelian projections. Therefore $\langle\mu^a\rangle\neq 0$ means Higgs
breaking of magnetic charge of type ``$a$'' in all of them.

Dual superconductivity is an absolute property, independent of the
abelian projection.

\begin{figure}[b]
\includegraphics*[width=\columnwidth]{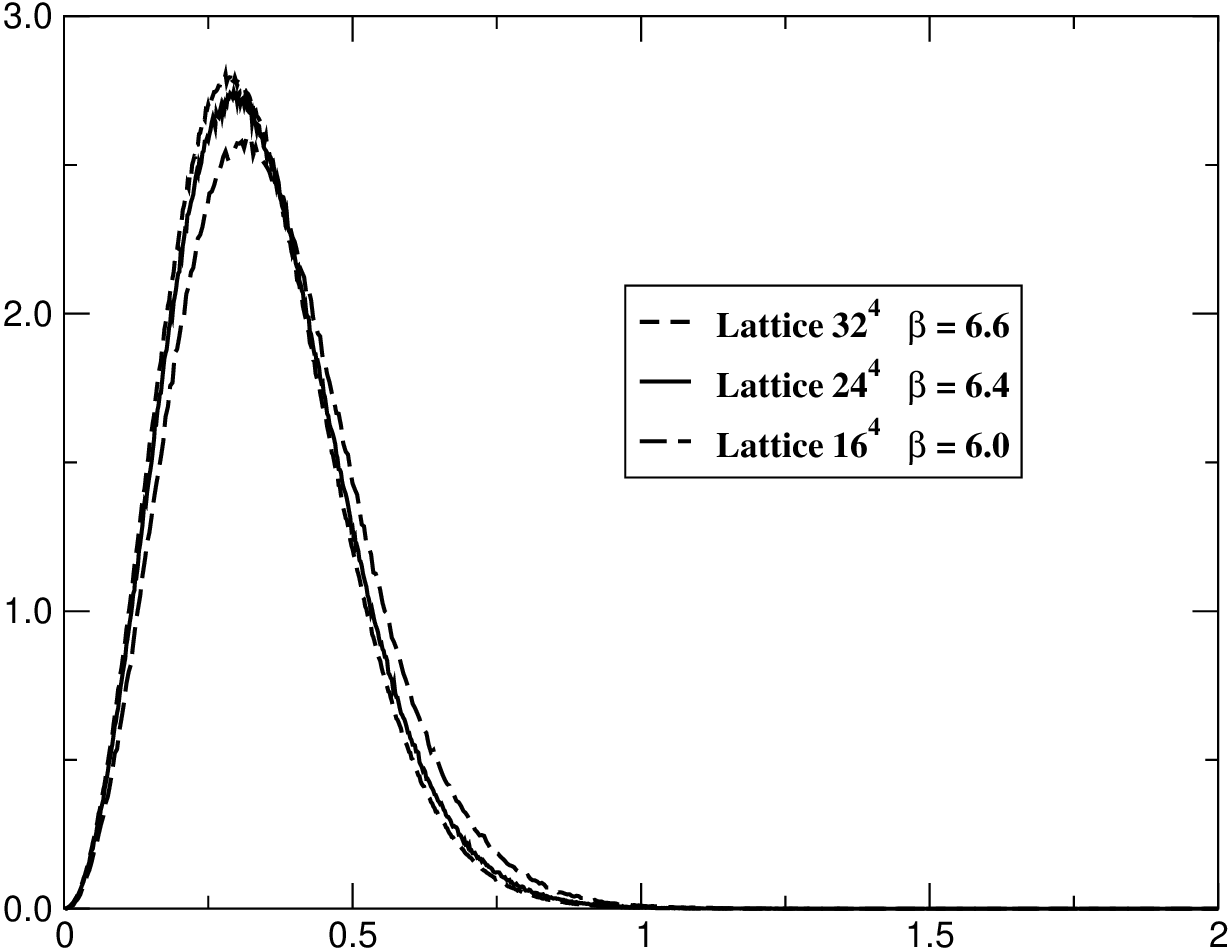}\\
\includegraphics*[width=\columnwidth]{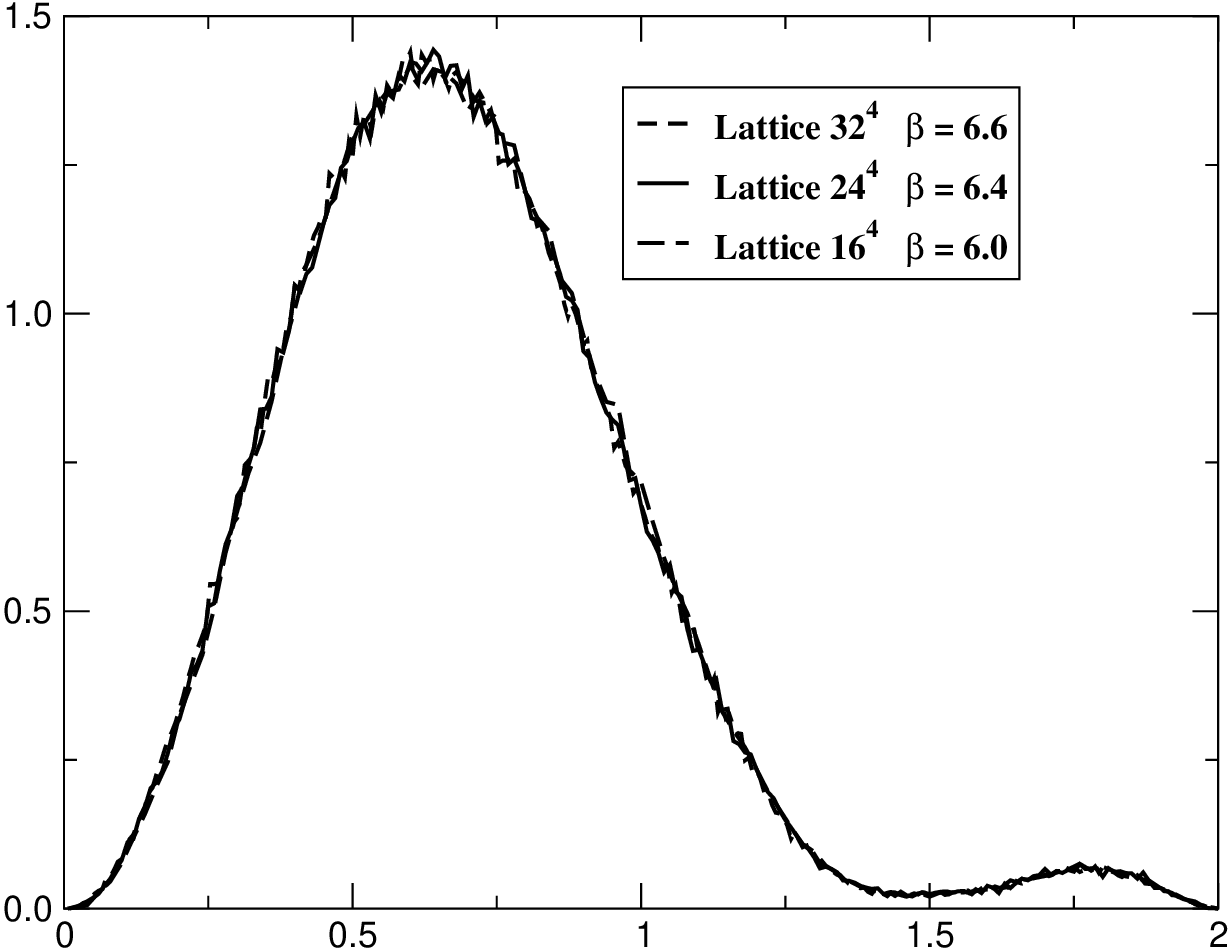}
\caption{Normalized distribution of the difference between the phases of consecutive eigenvalues (in units of $\pi$) of the plaquette operator (upper figure) and of the Polyakov loop (lower figure). The values $0$ or $2$ correspond to two coincident eigenvalues and thus to the presence of monopoles. The distribution has been measured at different values of $\beta$ and approximately equal physical volumes, in order to check the continuum limit.}\label{MONODEN}
\end{figure}

The density of monopoles is known to be
finite~\cite{Bornyakov:2003kn}. A direct test of that can be made by
looking at the distribution of the difference between eigenvalues of
any hermitian operator in the adjoint representation. Finding, in any
site of the lattice, a zero difference would mean that a monopole is
sitting on the site. The lattice can be made finer and finer by
increasing $\beta$. Fig.~\ref{MONODEN} shows that in no site the
eigenvalues coincide, and hence that the density of monopoles is
finite. The same happens at higher values of $\beta$.

A last point concerns the problem raised in Ref.~\cite{Frohlich:2000zp}
about the difficulty to obey Dirac condition for magnetic charges in
the presence of electric charges.  This is a general problem: in
$U(1)$ this difficulty shows up as the non existence of the duality
transformation, and this produces problems at short distances: an
operator like $\mu$ should be modified at distances $\mathcal{O}(a)$
by terms $\mathcal{O}(a^2)$. The authors of Ref.~\cite{Frohlich:2000zp}
suggest a modified formulation in which a coulombic magnetic field
produced by a monopole is replaced by a superposition of radial flux
tubes with quantized flux~\cite{Chernodub:2000wk}. This is a big
revision in the description of a monopole, but is in any case
unmanageable from the numerical point of view~\cite{Belavin:2002hb}.
In any case our operator is defined up to term $\mathcal{O}(a^2)$ as
explained in I Sect.~II, so that looking for $\mathcal{O}(a^2)$
corrections would be meaningless anyhow.  Contrary to $U(1)$
$\mathcal{O}(a^2)$ corrections in QCD should become unimportant in the
continuum limit because of asymptotic freedom. The problem is however
open and deserves further investigation.

\section{\label{SECT3}$N_f=2$ QCD}
The study of the QCD finite temperature transition for 2 flavors is of
fundamental importance.  The phase transition is well understood at
high masses ($m\geq2.5$ $GeV$), where the quarks decouple; the
transition is first order and the Polyakov line is a good order
parameter.  At $m\simeq 0$ a chiral transition exists, where chiral
symmetry is restored, and the chiral condensate
$\langle\bar\psi\psi\rangle$ is an order parameter.  At some
temperature also the axial $U_A(1)$ is expected to be restored: indeed
the topological susceptibility drops to zero around
$T_c$~\cite{Alles:2000cg}.  In principle at $m\simeq 0$ there are 3
transitions (chiral, axial $U_A(1)$, deconfinement): it is not clear
if they coincide.  An effective description of the chiral transition
can be given in terms of an effective free
energy~\cite{Pisarski:1983ms}. Assuming that the scalar and
pseudoscalar modes are the relevant critical degrees of freedom, there
is no infrared stable fixed points for $N_f\geq 3$ and the transition
is expected to be first order.  For $N_f=2$ the transition is first
order if the anomaly is negligible ($m_{\eta'}\approx 0$) at $T_c$; it
can be second order with symmetry $O(4)$ if the anomaly survives the
chiral transition.  In the first case the transition surface around
$m=0$ is first order.  If the chiral transition is instead second
order the surface is a crossover, and a tricritical point is expected
in the $\mu-T$ plane (see e.g.~\cite{Stephanov:1998dy}), detectable by
heavy ion experiments.

The first scenario is compatible with dual superconductivity of the
vacuum being the mechanism for confinement both in pure gauge and in
presence of dynamical quarks, independently of their mass: the
deconfinement transition is then always an order-disorder phase
transition, not a crossover.  If the second is the case instead, the
deconfining transition cannot be order-disorder, there is no order
parameter for confinement, and a state of a free quark can
continuously be transfered below the ``deconfining temperature'':
confined and deconfined then lose a definite meaning.

The question can be answered by a numerical study of the chiral phase
transition and has been the subject of extensive, even if not
conclusive,
literature~\cite{Fukugita:1990vu,Fukugita:1990dv,Brown:1990ev,Karsch:1993tv,Karsch:1994hm,Bernard:1999fv,AliKhan:2000iz}.
In Ref.~\cite{CHIPAPER} we have presented a large scale finite size
scaling analysis of the 2 flavor chiral transition by looking at the
specific heat, the chiral susceptibilty and the equation of state: we
have found clear disagreement with a second order $O(4)$ behaviour and
solid indications of the possible presence of a first order phase
transition. On the other hand the fact that dual superconductivity is
at work also in the presence of two dynamical flavours has been
already shown, for finite quark masses ($m_\pi/m_\rho \simeq 0.505$),
in Ref.~\cite{Carmona:2002yg}, using $\langle \mu \rangle $ as a
disorder parameter, and also in the Schr\"odinger functional approach
in Ref.~\cite{Cea:2004ux}.

In this Section we present a finite size scaling analysis of $\langle
\mu \rangle $ around the chiral phase transition: if $\langle \mu
\rangle $ is the correct (dis)order parameter for confinement, this
analysis will give the correct critical indexes.

The parameter $\langle \mu \rangle$ is defined on the lattice as (see
I and II)
\begin{eqnarray}
\label{defmu}
\langle \mu \rangle = \frac{\tilde{Z}}{Z} \; ,\nonumber \\
Z = \int \left( {\cal D}U \right)  e^{-\beta S} \; ,\nonumber \\   
\tilde{Z} = \int \left( {\cal D}U \right)  e^{-\beta \tilde{S}} \; .
\end{eqnarray}
$\tilde{Z}$ is obtained from $Z$ by changing the action in 
the time slice $x_0$, $S \to \tilde{S} = S + \Delta S$.
In the Abelian projected gauge the plaquettes
\begin{eqnarray}
&&\Pi_{i0} (\vec{x},x_0) = \nonumber \\
&& = U_i (\vec{x},x_0) U_0 (\vec{x} + \hat{\imath},x_0)
U_i^\dagger(\vec{x},x_0 + \hat{0}) U_0^\dagger (\vec{x},x_0)
\end{eqnarray}
are changed by substituting
\begin{eqnarray}
U_i(\vec{x},x_0) \to \tilde{U}_i(\vec{x},x_0) \equiv
U_i(\vec{x},x_0) e^{i T b^i_\perp (\vec{x} - \vec{y})}
\end{eqnarray}
where $T$ is the diagonal gauge group
generator corresponding to the monopole species chosen.  The numerical
determination of $\langle \mu \rangle$ is very difficult, since
$\langle \mu \rangle$ is expressed as the ratio of two partition
functions.  Instead of $\langle \mu \rangle$ we measure the quantity
\begin{eqnarray}
\rho = \frac{d}{d \beta} \ln \langle \mu \rangle \; .
\end{eqnarray}
It follows from Eq. (\ref{defmu}) that
\begin{eqnarray}
\rho = \langle S \rangle_S -  \langle \tilde{S} \rangle_{\tilde{S}} \; ,
\label{rhoferm}
\end{eqnarray}
the subscript meaning the action by which the average is performed. 
In terms of $\rho$
\begin{eqnarray} 
\label{mufromrho}
\langle \mu \rangle = \exp\left(\int_0^{\beta} \rho(\beta^{\prime})\mbox{d}\beta^{\prime}\right) \; .
\end{eqnarray}
A drop of $\langle \mu \rangle$ at the phase transition corresponds to
a strong negative peak of $\rho$. A negative value of $\rho$ diverging
in the thermodynamical limit in the deconfined phase means $\langle
\mu
\rangle$ being exactly zero on that side (see I, II, III).

We have made simulations with two degenerate flavors of Kogut-Susskind
quarks, using the standard gauge and fermion actions.  Configuration
updating was performed using the standard Hybrid R
algorithm~\cite{Gottlieb:1987mq}. The lattice temporal size was fixed
at $N_t=4$.  Different spatial sizes ($L = 12,16,20,24,32$) and values
of the quark mass were used. For a more detailed account on simulation
parameters we refer to~\cite{CHIPAPER}.

\begin{figure}[t!]
\includegraphics*[height=5.8cm,width=\columnwidth]{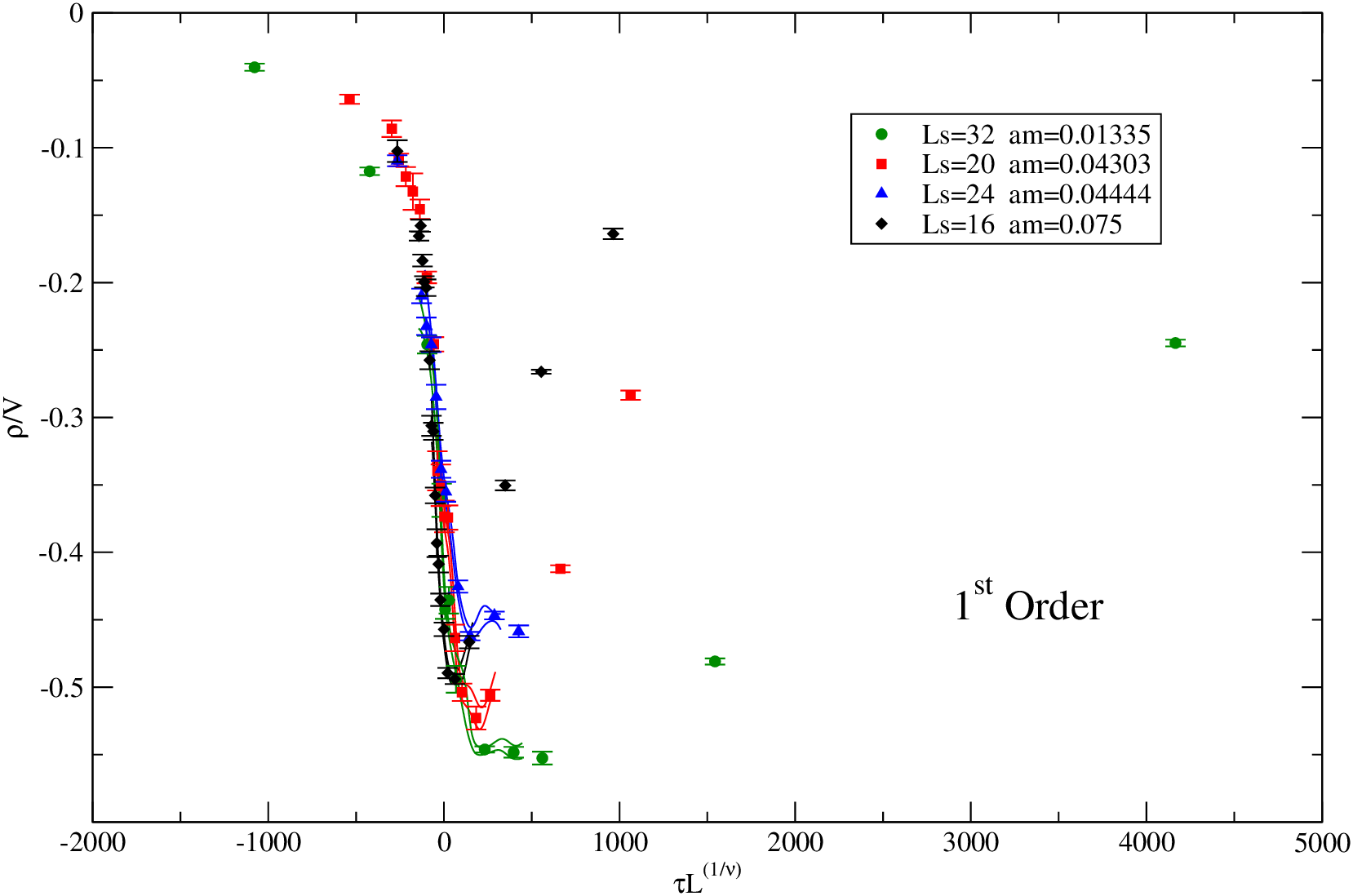}
\caption{Finite size scaling of $\rho$ according to the critical indexes of first order transition.}\label{fig1st} 
\end{figure}

\begin{figure}[thb]
\includegraphics*[height=5.8cm,width=\columnwidth]{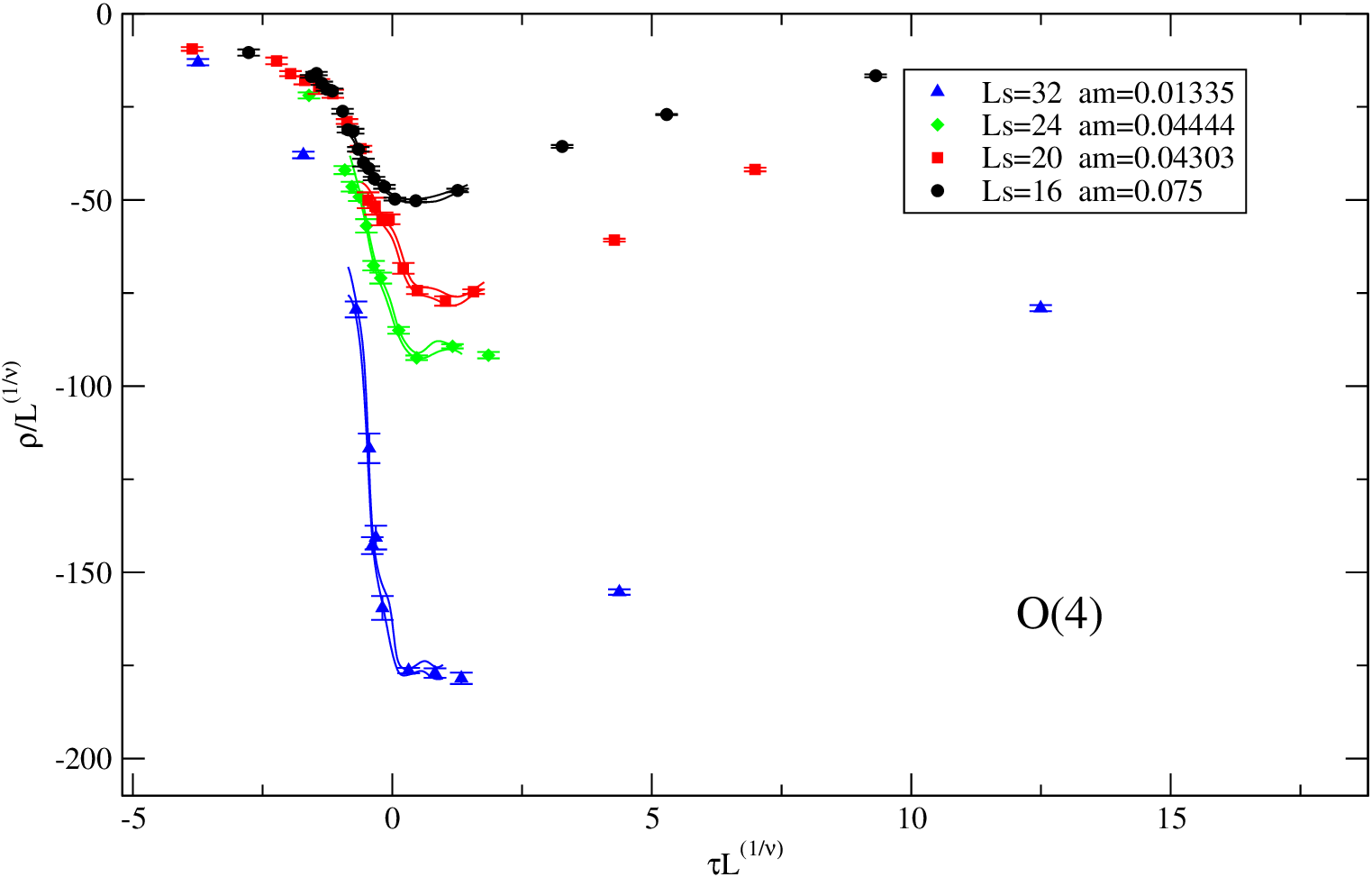}
\caption{Finite size scaling of $\rho$ according to the $O(4)$ 
critical indexes.}\label{figO4} 
\end{figure}

We can assume the following general scaling form for $\langle \mu
\rangle$ around the phase transition:
\begin{eqnarray}
\langle \mu \rangle = L^k \Phi ( \tau L^{1/\nu},m L^{y_h} ) \nonumber
\end{eqnarray}
where $\tau$ is the reduced temperature and $m$ the quark mass.
Sending $L$ to infinity keeping $\xi/L$ or $\tau L^{1/\nu}$ fixed,
must be a convergent limit. Analyticity
arguments~\cite{D'Elia:2004xi,Karsch:1994hm} suggest that in the
infinite volume limit the mass dependence in the scaling function
factorizes, so that $\rho = \frac{d}{d \beta} \ln \langle \mu \rangle$
does not depend on the mass. In fact, the dependence on $m L^{y_h}$
must then cancel the dependence on the factor $L^k$ in front.  We then
obtain the following scaling law:
\begin{eqnarray} 
\rho = L^{1/\nu} \phi(\tau L^{1/\nu}) \; .
\end{eqnarray}
the same as in the quenched case.

In Figure 2 we show the quality of scaling assuming $\nu = 1/3$,
i.e. a first order phase transition: a good agreement is clearly
visible.  The deviations from scaling in the deconfined region are
expected~\cite{D'Elia:2003xn} and are related to the disorder
parameter $\langle \mu \rangle$ being exactly zero on that side.

For comparison we show in Figure 3 the quality of scaling assuming the
$O(4)$ critical index $\nu = 0.75$.
	
The $O(4)$ universality class is clearly excluded, while there is good
agreement with a first order phase transition. This confirms results
obtained through an analysis of the specific heat and of the chiral
susceptibility~\cite{CHIPAPER}.

\section{\label{SECT4}Conclusions}

QCD with $N_f=2$ is a specially interesting system to investigate
confinement. The order of the chiral phase transition can indeed
decide if the deconfining transition is a real transition, as required
by a mechanism of confinement like dual superconductivity, or a
trivial cross-over.

More generally the very possibility of defining confined versus
deconfined relies on the existence of an order parameter. This
parameter exists for the quenched theory, where the transition is
first order, but is not defined for full QCD. For full QCD only a
chiral order parameter $\langle\bar\psi\psi\rangle$ is defined, and
only at low enough quark masses.

We have shown that an order parameter $\langle\mu\rangle$ can be
defined, which, independently of the presence of quarks and of their
masses, characterizes confinement and deconfinement.
$\langle\mu\rangle\neq 0$ at $T<T_c$; $\langle\mu\rangle$ strictly
zero in the limit $V\rightarrow\infty$ for $T>T_c$.

A finite size scaling analysis gives the correct (pseudo)critical
indexes both for the quenched and for the $N_f=2$ case. For $N_f=2$
the deconfining transition as seen by looking at $\langle\mu\rangle$
is first order.  This is the main result of the present article. The
same indication comes from the analysis of the specific
heat~\cite{CHIPAPER}, which is independent of any prejudice on the
mechanism of confinement.

Further study is on the way with improved actions and algorithm to
settle this issue, as well as to study the behavior of $m_{\eta'}$ at
the chiral transition point.

A consistent picture starts to emerge for confinement. A theoretical
problem in the definition of $\langle\mu\rangle$ in the presence of
electric charges exists, which deserves further study.

\begin{acknowledgments}
Discussions with J.~M.~Carmona and L.~Del~Debbio in the initial stages
of this work are acknowledged.
\end{acknowledgments}

\end{document}